\documentclass[%
 aps,
 prb,
 amsmath,amssymb,
 showpacs,
 reprint,%
]{revtex4-1}

\usepackage{graphicx}% Include figure files
\usepackage{dcolumn}% Align table columns on decimal point
\usepackage{bm}% bold math
\usepackage{verbatim}
\begin{document}

\title{Periodically driven small polarons}
%\thanks{Footnote to title of article.}

\author{P.E. Kornilovitch}
 \email{pavel.kornilovich@gmail.com}
 \affiliation{Department of Physics, Oregon State University, Corvallis, Oregon 97331, USA} 

\date{\today}  % It is always \today, but any date may be explicitly specified

\begin{abstract}

Small lattice polarons driven by strong external electric fields are considered. A time-dependent Schr\"odinger equation is integrated directly in time domain. The field agitates ions both directly and through modulation of carrier density. It is found that when the field is in resonance with local ion oscillations, the polaron is liberated from its self-induced trap and the tunneling frequency increases exponentially.  

\end{abstract}

\pacs{33.80.-b, 63.20.-e}     % PACS, the Physics and Astronomy Classification Scheme

%\keywords{???}               % Use showkeys class option if keyword display desired

\maketitle

\section{\label{drivpol:sec:one}
Introduction
}

Polaron is a carrier in a condensed-matter system that deforms surrounding atoms or ions from their equilibrium positions. A stable polaron moves through the system together with the deformation. The deformation impedes polaron motion, increases its effective mass and reduces mobility. In polar solids with low carrier density, the carrier-ion interaction derives from an unscreened Coulomb interaction which is strong at short distances. Polaron transport properties sharply, often exponentially, depend on ion positions, and are a sensitive probe of the ion subsystem. Traditionally, ion motion has been controlled through global variables such as temperature or pressure, or through local modifications such as isotope substitutions. Recently, however, direct excitation of ions with lasers became possible.~\cite{Hu2014} Ions are driven into resonance by strong laser fields and acquire oscillation amplitudes as large as several percent of a lattice constant.~\cite{Hu2014} Such displacements significantly reduce or even completely eliminate the deformation caused by the interaction with a carrier. As a result, the carrier is ``liberated'' from the potential trap created by the ions and its transport changes dramatically. External driving of ions becomes therefore a powerful tool to probe intrinsic polaron properties.      

Polaron physics has a rich history that dates back to Landau,~\cite{Landau1933} Pekar,~\cite{Pekar1946,Landau1948} and Fr\"ohlich.~\cite{Froehlich1954} The development of the field is well documented in several books and reviews.~\cite{Pekar1951,Kuper1963,Appel1968,Devreese1972,Levinson1973,Firsov1975,Devreese1984,Alexandrov1996,Devreese1996,Alexandrov2007,Devreese2009,Emin2013} In addition, for decades the polaron has served as a testing ground for novel analytical and numerical techniques including perturbative expansions,~\cite{Roeseler1968,Miyake1975,Smondyrev1986,Stephan1996,Alexandrov2003} path integrals,~\cite{Feynman1955,Feynman1962,Verbist1991} Monte Carlo methods,~\cite{Hirsch1982,Fradkin1983,DeRaedt1982,Alexandrou1990,Alexandrou1992,McKenzie1996,Kornilovitch1997,Kornilovitch1998,Prokofev1998,Mishchenko2000,Mishchenko2003,Macridin2004,Hague2007} cluster diagonalization,~\cite{Ranninger1992,Kabanov1994,Marsigio1995,Wellein1996,Wellein1997} variational,~\cite{Feynman1955,Lee1953,Romero1998,Bonca1999,Bonca2000,Chakrabarti2007,Chakraborty2011,Chandler2016} and other~\cite{Tjablikov1951,Sewell1958,Holstein1959a,Holstein1959b,Friedman1963,Eagles1963,Lang1963,Emin1969,Emin1976,Kabanov1993,Freericks1993,Ciuchi1997,Zhang1998,Jeckelmann1998,Hohenadler2004,Berciu2006,Goodvin2006,Fratini2006} methods. However, there has hardly been any work on externally driven polarons or bipolarons. Most theoretical treatments of polaron-light interactions are based on the linear response theory,~\cite{Feynman1962,Lang1963,Eagles1963,Friedman1963,Devreese1972,Mishchenko2003,Fratini2006,Devreese2009} and as such assume weak coupling between a charge carrier and an external electric field. The field is treated as a perturbation that does not change polaron states but only causes transitions between them. In this work, we are interested in the opposite limit of strong coupling when the field directly alters the local state of the polaron. Arguably, this is a much harder mathematical problem, and this is perhaps one reason why it has not been widely addressed before. If fact, we are not aware of any publication on this subject prior to our own recent paper.~\cite{Kornilovitch2016}

\begin{figure}[t]
\includegraphics[width=0.48\textwidth]{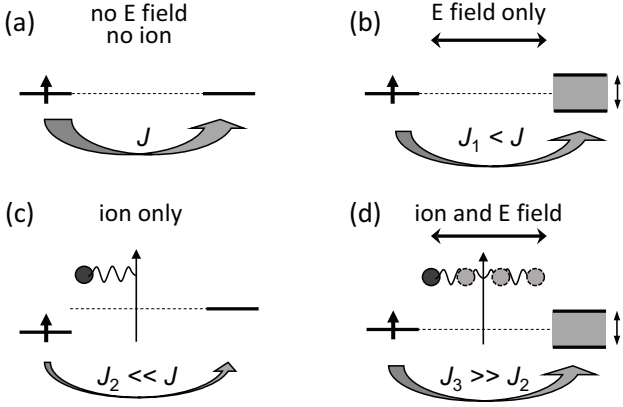}
\caption{The effects of three interaction types on carrier tunneling. (a) A free carrier on two sites. (b) An external field modulates on-site energies and reduces the tunneling rate. (c) Carrier-ion interaction skews the potential and impedes tunneling (polaron formation). (d) The field drives the ion in resonance symmetrizing the potential and {\em increasing} the tunneling rate relative to case (c).}
\label{drivpol:fig:one}
\end{figure}

The basic physics of driven polarons can be understood on a two-site model system. In the absence of both ion displacement and electric field, Fig.~\ref{drivpol:fig:one}(a), two atomic levels are in resonance, leading to carrier transfer with a bare amplitude $J$. An external electric field, Fig.~\ref{drivpol:fig:one}(b), periodically drives the levels on and off resonance, which on average reduces the tunneling amplitude to $J_1 < J$. When an ion is included in the system, Fig.~\ref{drivpol:fig:one}(c), its displacement drives the levels off resonance on a more sustained basis. Tunneling is possible only with simultaneous reversal of the deformation, and as such is exponentially suppressed. This constitutes polaron formation. However, a sufficiently strong external field can drive the ion in resonance, see Fig.~\ref{drivpol:fig:one}(d). That symmetrizes the potential so that the two levels come into resonance periodically. As a result,~\cite{Kornilovitch2016} the tunneling amplitude increases {\em exponentially} compared to the undriven case, $J_3 \gg J_2$, but of course remains smaller than bare amplitude $J$.     

The goal of the present work is to extend the treatment of Ref.~[\onlinecite{Kornilovitch2016}] beyond the adiabatic approximation. Additionally, indirect ion-field interaction through modulation of carrier density is also included in the picture. In general, the direct and indirect interactions compete with each other and their balance should be carefully analyzed. Only single polaron case is studied while the bipolaron case is left for future work.

\section{\label{drivpol:sec:two}
Two-site driven polaron model
}

We consider one carrier of charge $z_c$ moving between two tight-binding sites $\vert 1 \rangle$ and $\vert 2 \rangle$ with a hopping integral $J$. The distance between the sites is $2b$. One ion of mass $M$, frequency $\Omega$, charge $z_i$, and dynamic coordinate $y$ is positioned symmetrically between the sites. The symmetric position and polarization of ion displacement (parallel to the line between the sites) is inspired by the layered structure of high-temperature superconductors. Note that this arrangement is different from the Holstein model of molecular crystals,~\cite{Holstein1959a} where each ion interacts with one lattice site only. The full Hamiltoni\-an consists of five terms: free carrier $H_c$, free ion $H_i$, carrier-ion $H_{ci}$, carrier-field $H_{cf}$, and ion-field $H_{if}$:
\begin{eqnarray}
H      & = & H_{c} + H_{i} + H_{ci} + H_{cf} + H_{if}  ,
\label{drivpol:eq:one}    \\
H_{c}  & = & - J \left( \vert 1 \rangle \langle 2 \vert + \vert 2 \rangle \langle 1\vert \right)   ,
\label{drivpol:eq:two}    \\
H_{i}  & = & - \frac{\hbar^2}{2M} \frac{\partial^2}{\partial y^2} + \frac{1}{2} \, M \Omega^2 y^2  , 
\label{drivpol:eq:three}  \\
H_{ci} & = & - g \, y  
\left( \vert 1 \rangle \langle 1 \vert - \vert 2 \rangle \langle 2 \vert \right) ,
\label{drivpol:eq:four}   \\
H_{cf} & = & z_c \vert e \vert {\cal E} \, b  
\left( \vert 1 \rangle \langle 1 \vert - \vert 2 \rangle \langle 2 \vert \right) \, \sin{\omega_0 t} \: ,
\label{drivpol:eq:five}   \\
H_{if} & = & - z_i \vert e \vert {\cal E} \, y \, \sin{\omega_0 t} \: .
\label{drivpol:eq:six}
\end{eqnarray}
Here ${\cal E}$ and $\omega_0$ are the field's amplitude and frequency, and $g$ is the {\em force} between the ion and the carrier on either site. The force is taken to be independent of $y$, i.e., carrier-ion interaction is treated in the linear approximation. Note that Eqs.~(\ref{drivpol:eq:one})-(\ref{drivpol:eq:six}) are invariant under simultaneous site inversion $\vert 1 \rangle \leftrightarrow \vert 2 \rangle$, coordinate inversion $y \rightarrow -y$ and time inversion $t \rightarrow -t$.

Because of a large number of physical parameters, it is useful to transform to dimensionless variables. It is convenient to measure everything in oscillator units. We choose $\Omega^{-1}$, $\hbar\Omega$, and $y_0 = \sqrt{\hbar/(M\Omega)}$ as the units of time, energy, and length, respectively. Accordingly, dimensionless time $\tau$, coordinate $\zeta$, frequency $\omega$, and hopping integral $j$ are introduced as follows:
\begin{equation}
\tau = \Omega t \, , \hspace{0.5cm} \zeta = \frac{y}{y_0} \, , \hspace{0.5cm}
\omega = \frac{\omega_0}{\Omega} \, , \hspace{0.5cm} j = \frac{J}{\hbar \Omega} \: .
\label{drivpol:eq:seven}
\end{equation}
Dimensionless coupling constants between the field and the carrier and between the field and the ion follow from the transformation:      
\begin{equation}
\alpha_c = \frac{\vert e \vert {\cal E} b}{\hbar\Omega} \, , \hspace{0.5cm} 
\alpha_i = \frac{\vert e \vert {\cal E}}{\hbar\Omega} \sqrt{\frac{\hbar}{M\Omega}} \: .
\label{drivpol:eq:eight}
\end{equation}
To estimate $\alpha_c$ and $\alpha_i$, one can use values reported by Hu et al~\cite{Hu2014} in their studies of dynamically stabilized superconductivity in YBa$_2$Cu$_3$O$_{6.5}$: ${\cal E} = 3.0$ MV cm$^{-1}$ and $\hbar\Omega = 83$ meV. Using the distance between copper-oxygen bilayers (8.2 \AA) for $2b$ and the mass and charge of the oxygen ion, one obtains $\alpha_c \approx 1.5$ and $\alpha_i \approx 0.02$. Despite the fact that $\alpha_i \ll \alpha_c$, effects of the two coupling types are comparable, as will be detailed below.  

The carrier-ion coupling constant $\lambda$ is now discussed. Commonly,~\cite{Alexandrov1996} $\lambda$ is defined as the ratio of {\em polaron shift} $E_p = g^2/(2M\Omega^2)$ (polaron energy in the atomic limit $J = 0$) to half of bare bandwidth $D$ (i.e. the lowest energy of a free carrier). The half bandwidth in a {\em two-site} system is $D_2 = J$. Accordingly, $\lambda$ is defined here as 
\begin{equation}
\lambda = \frac{g^2}{2 M \Omega^2 J} \: .
\label{drivpol:eq:nine}
\end{equation}
Expression~(\ref{drivpol:eq:nine}) is quadratic in $g$, so the same $\lambda$ describes both attraction and repulsion. To distinguish between the two possibilities, a sign variable $z_{ci}$ is introduced
\begin{equation}
z_{ci} = \frac{z_c}{\vert z_c \vert} \cdot \frac{z_i}{\vert z_i \vert}\: .
\label{drivpol:eq:nineone}
\end{equation}
Accordingly, $g$ in Eq.~(\ref{drivpol:eq:four}) can be written as:  
\begin{equation}
g = z_{ci} \sqrt{2 M \Omega^2 J \lambda} \: .
\label{drivpol:eq:ten}
\end{equation}
For the sake of visual clarity and keeping in mind the physics of YB$_2$C$_3$O$_{6.5}$, hereafter we will focus on the case of attraction, i.e. positive $z_c = + 1$ (a hole inside a copper-oxygen plane) and negative $z_i = -2$ (an apical oxygen ion), which implies $g < 0$. With such a choice, the on-site energy of $\vert 1 \rangle$ decreases for negative $y$, see Eq.~(\ref{drivpol:eq:four}).   

The full wave function is a two-element array $\{ \psi_{1} ; \psi_{2} \}$, both elements being functions of time $\tau$ and ion displacement $\zeta$. Collecting all the definitions, the Schr\"odinger equation reads 
\begin{equation}
i \frac{\partial}{\partial \tau} \left( 
\begin{array}{c} \psi_1 \\ \psi_2 \end{array} \right) =
\left( \begin{array}{cc} \hat{h}_{1\zeta} & - j              \\ 
                           - j            & \hat{h}_{2\zeta}  
       \end{array}  \right)
\left( \begin{array}{c} \psi_1 \\ \psi_2 \end{array} \right) ,
\label{drivpol:eq:eleven}       
\end{equation}
where      
\begin{eqnarray}
\hat{h}_{1\zeta,2\zeta} & = &   
- \frac{1}{2} \frac{\partial^2}{\partial \zeta^2} + \frac{1}{2} \, \zeta^2 
\mp z_{ci} \sqrt{2 j \lambda} \cdot \zeta  
\nonumber \\
& & - \left( z_i \alpha_i \cdot \zeta \mp z_c \alpha_c \right) \sin{(\omega \tau)} \: .
\label{drivpol:eq:twelve}
\end{eqnarray}
Normalization is chosen to be
\begin{equation}
\int^{\infty}_{-\infty}  
\left\{ \vert \psi_1(\tau,\zeta) \vert^2 + \vert \psi_2(\tau,\zeta) \vert^2 \right\} d\zeta = 1 \: .
\label{drivpol:eq:thirteen}
\end{equation}

\section{\label{drivpol:sec:three}
Numerical method
}

Equation (\ref{drivpol:eq:eleven}) does not admit analytical solutions and has to be integrated numerically. One approach is based on Floquet theory and spectral expansion of $\psi(\tau)$ in a Floquet basis.~\cite{Grossmann1991,Grifoni1998} Here we adopt an alternative approach of direct integration in time domain. The particular differencing scheme is based on the Cayley form of the finite-time evolution operator~\cite{Goldberg1967,Press1992}   
\begin{equation}
e^{- i ( \hat{h}_{\zeta} + \hat{h}_{j} ) \triangle \tau } = 
\frac{ \left( 1 - \frac{i}{2} \hat{h}_{\zeta} \triangle \tau \right) 
       \left( 1 - \frac{i}{2} \hat{h}_{j}     \triangle \tau \right) }
     { \left( 1 + \frac{i}{2} \hat{h}_{\zeta} \triangle \tau \right) 
       \left( 1 + \frac{i}{2} \hat{h}_{j}     \triangle \tau \right) }   +
O( \triangle \tau^3 ) \: ,
\label{drivpol:eq:fourteen}
\end{equation}
where 
\begin{equation}
\hat{h}_{\zeta} = \left( \begin{array}{cc} 
\hat{h}_{1\zeta} &  0                 \\ 
      0          & \hat{h}_{2\zeta}  \end{array} \right)  , 
\hspace{0.5cm}
\hat{h}_{j} = \left( \begin{array}{cc} 
      0   &  -j                 \\ 
     -j   &   0  \end{array} \right)  .
\label{drivpol:eq:fifteen}
\end{equation}
This representation is unitary and accurate to order $(\triangle \tau)^2$ even for non-commuting operators $\hat{h}_{\zeta}$ and $\hat{h}_{j}$, which is the case here. Equation~(\ref{drivpol:eq:fourteen}) leads to an unconditionally stable implicit time-stepping rule~\cite{Crank1947}
\begin{eqnarray}
\left( 1 + \frac{i}{2} \hat{h}_{\zeta} \triangle \tau \right) 
\left( 1 + \frac{i}{2} \hat{h}_{j}     \triangle \tau \right)  
\left\{ \begin{array}{c}
\psi_1( \tau_{k+1} , \zeta ) \\ \psi_2( \tau_{k+1} , \zeta ) 
\end{array} \right\} =  & & 
\nonumber \\
\left( 1 - \frac{i}{2} \hat{h}_{\zeta} \triangle \tau \right) 
\left( 1 - \frac{i}{2} \hat{h}_{j}     \triangle \tau \right)  
\left\{ \begin{array}{c}
\psi_1( \tau_{k} , \zeta ) \\ \psi_2( \tau_{k} , \zeta ) 
\end{array} \right\} .   & &   
\label{drivpol:eq:sixteen}
\end{eqnarray}
where $\tau_k$ is the $k$-th time step. Factorization of the evolution operator into $\zeta$ and $j$ parts enables sequential inversion: first along $\zeta$ axis and then along the site index. The time-stepping procedure is as follows. (i) Both wave function components are discretized along $\zeta$ dimension: $\psi_{1,2}(\tau_k,\zeta) \rightarrow \psi_{1,2}(\tau_k,\zeta_l)$. (ii) Starting with $\psi_{1,2}(\tau_k,\zeta_l)$, the right hand side of Eq.~(\ref{drivpol:eq:sixteen}) is computed. First, $j$ part is applied which amounts to a $(2 \times 2)$ matrix multiplication for each $l$. Second, $\zeta$ part is applied, which is performed independently for $\psi_1$ and $\psi_2$ [$\hat{h}_{\zeta}$ in Eq.~(\ref{drivpol:eq:fifteen}) is diagonal]. The resulting right hand side is a two-row matrix $\{ b^{l}_{1} ; b^{l}_{2} \}$. (iii) The left hand side of Eq.~(\ref{drivpol:eq:sixteen}) is inverted, again in two consecutive steps. First, an intermediate function $\psi^{\ast}$ is found from the following relation
\begin{equation}
\left( 1 + \frac{i}{2} \hat{h}_{\zeta} \triangle \tau \right) 
\left\{ \begin{array}{c}
\psi^{\ast}_1( \zeta_l ) \\ \psi^{\ast}_2( \zeta_l ) 
\end{array} \right\}   = 
\left\{ \begin{array}{c}
b^{l}_1 \\ b^{l}_2  
\end{array} \right\}  .
\label{drivpol:eq:seventeen}
\end{equation}
Since $\hat{h}_{\zeta}$ is diagonal, $\psi^{\ast}_1$ and $\psi^{\ast}_2$ are computed independently using the method of Ref.~[\onlinecite{Goldberg1967}]. (iv) Finally, $\psi(\tau_{k+1},\zeta_l)$ is found by solving
\begin{equation}
\left( 1 + \frac{i}{2} \hat{h}_{j} \triangle \tau \right) 
\left\{ \begin{array}{c}
\psi_1( \tau_{k+1} , \zeta_l ) \\ \psi_2( \tau_{k+1} , \zeta_l ) 
\end{array} \right\}   = 
\left\{ \begin{array}{c}
\psi^{\ast}_1( \zeta_l ) \\ \psi^{\ast}_2( \zeta_l ) 
\end{array} \right\}  .
\label{drivpol:eq:eighteen}
\end{equation}
Since operator $\hat{h}_j$ does not mix spatial coordinates, solving the last equation amounts to inverting a $(2 \times 2)$ matrix for each $l$. (v) The above sequence is repeated for the entire time interval of interest. Most results presented in this paper were obtained with $\triangle \tau = 0.001$ and $\triangle \zeta = 0.1$. Thus, time evolution over a total time of $\tau_{\rm max} = 10^4$ requires $10^7$ time steps.   
   
Choice of initial conditions $\psi_{10,20}$ is now discussed. The primary quantity of interest in this work is pola\-ron tunneling frequency between the sites. Therefore, we seek an initial state that is localized at one of the two sites. Adiabatic approximation provides a good starting point. Remove the second derivative and assume $\alpha_c = \alpha_e = 0$ in Eq.~(\ref{drivpol:eq:twelve}). Then fix $\zeta$ and solve the two-site one-carrier problem to obtain the polaron adiabatic potential
\begin{equation}
w(\zeta) = \frac{1}{2} \, \zeta^2 - \sqrt{ j^2 + 2 \lambda j \cdot \zeta^2 } \: . 
\label{drivpol:eq:nineteen}
\end{equation}
At $\lambda > \lambda_{\rm cr} = \frac{1}{2}$, $w(\zeta)$ develops two symmetric minima at 
\begin{equation}
\zeta_0 = \pm \sqrt{ \frac{2 j}{\lambda} \left( \lambda^2 - \lambda^2_{\rm cr} \right) } \: . 
\label{drivpol:eq:twenty}
\end{equation}
Near $\zeta_0$, the potential is quadratic with renormalized frequency (in physical units)
\begin{equation}
\tilde{\Omega} = \Omega \sqrt{ 1 - \frac{\lambda^2_{\rm cr}}{\lambda^2} } 
\equiv \Omega \cdot \tilde{\omega} \: . 
\label{drivpol:eq:twentyone}
\end{equation}
Accordingly, a good starting wave function describing the polaron in one of its minima is the ground state of an oscillator with frequency $\tilde{\omega}$ shifted by $\zeta_0$ 
\begin{equation}
\psi_{10,20} = \kappa_{1,2} \cdot \frac{\tilde{\omega}^{1/4}}{\pi^{1/4}} \, 
e^{ - \frac{1}{2} \tilde{\omega} ( \zeta - \zeta_0 )^2  } \: . 
\label{drivpol:eq:twentytwo}
\end{equation}
Weights $\kappa_{1,2}$ follow from solving the two-site problem
\begin{equation}
\frac{\kappa_2}{\kappa_1} = 2 \lambda - 2 \sqrt{ \lambda^2 - \lambda^2_{\rm cr} } \: , 
\label{drivpol:eq:twentythree}
\end{equation}
and from the normalization condition $\vert \kappa_1 \vert^2 + \vert \kappa_2 \vert^2 = 1$. 

The initial wave function can be further improved by applying the projection operator 
\begin{equation}
\psi_{0\beta} = e^{ - \beta ( \hat{h}_{\zeta} + \hat{h}_j ) } \psi_0 \: , 
\label{drivpol:eq:twentyfour}
\end{equation}
where $\beta$ is a positive dimensionless number. Calculation of Eq.~(\ref{drivpol:eq:twentyfour}) is also performed in a step-like fashion, by splitting total imaginary time $\beta$ into small steps $\triangle \beta \ll 1$ and replacing the short-time evolution operator with a $(\triangle \beta)^2$-accurate representation~\cite{DeRaedt1987} 
\begin{eqnarray}
& & e^{ - (\triangle \beta) ( \hat{h}_{\zeta} + \hat{h}_j ) } \approx 
\nonumber \\
& & \left( 1 - \frac{1}{2} \hat{h}_{j}     \triangle \beta \right) 
    \left( 1 -             \hat{h}_{\zeta} \triangle \beta \right) 
    \left( 1 - \frac{1}{2} \hat{h}_{j}     \triangle \beta \right) . 
\label{drivpol:eq:twentyfive}
\end{eqnarray}
Application of all the factors is done in the same way as in the case of real-time evolution, as described above. Since the projection operator does not conserve normalization, $\psi_0$ is normalized after every time step $\triangle \beta$. Most results presented below were obtained with $\beta = 2.0$ and $\triangle \beta = 0.001$.

\begin{figure*}[t]
\includegraphics[width=0.95\textwidth]{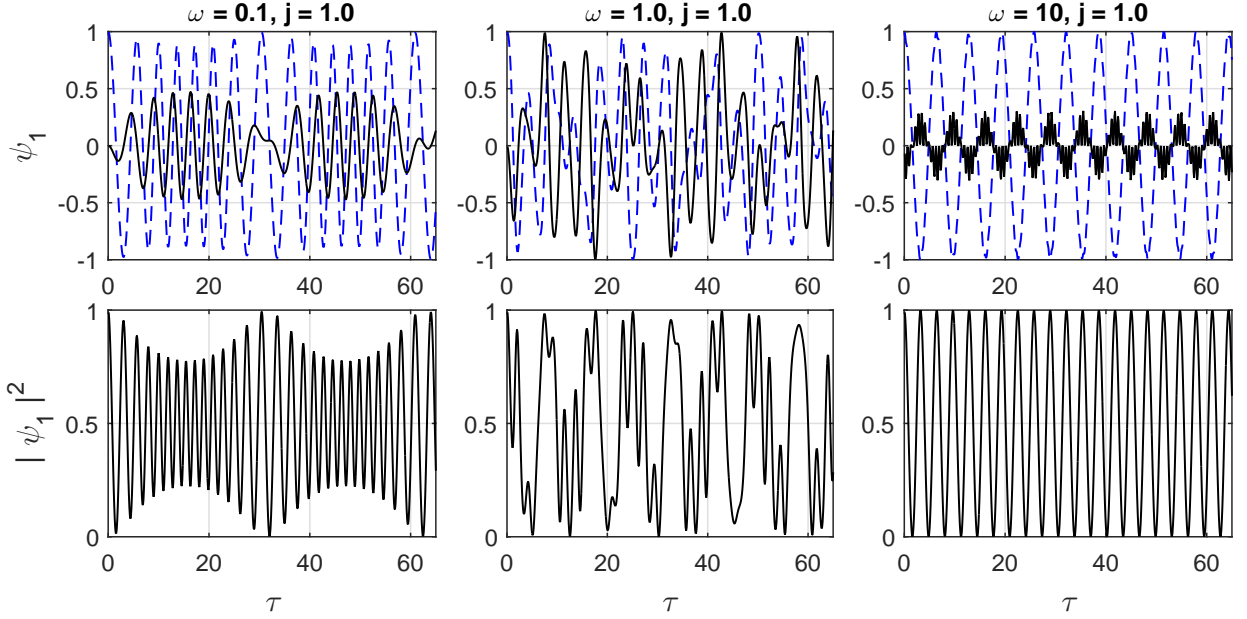}
\caption{(Color online) Numerical solution of Eqs.~(\ref{drivpol:eq:twentyseven}) and (\ref{drivpol:eq:twentyeight}) for $j = 1.0$, $\alpha_c = 1.5$ and initial conditions $\psi_1(0) = 1$ and $\psi_2(0) = 0$. The top row is wave function $\psi_1(\tau)$, both real part (dashed line) and imaginary part (solid line). The bottom row is the probability to reside on site 1. }
\label{drivpol:fig:two}
\end{figure*}

For the purposes of this paper, a full wave function contains too much information. To visualize tunneling rate, it is more convenient to consider an integral probability for the carrier to occupy either site 1 or site 2
\begin{equation}
P_{1,2}(\tau) = \int^{\infty}_{-\infty} d\zeta \left\vert \psi_{1,2}(\tau,\zeta) \right\vert^2 \: , 
\label{drivpol:eq:twentyfiveone}
\end{equation}
or probability for the ion to be on the left (L) or on the right (R) from the symmetry point $\zeta = 0$
\begin{equation}
P_{(L,R)}(\tau) = \int^{(0,\infty)}_{(-\infty,0)}   
\left\{ \vert \psi_1(\tau,\zeta) \vert^2 + \vert \psi_2(\tau,\zeta) \vert^2 \right\} d\zeta \: .
\label{drivpol:eq:twentyfivetwo}
\end{equation}
Another quantity of interest is instantaneous polaron energy without ion-field and carrier-field contributions
\begin{eqnarray}
E(\tau) & = & \langle \Psi(\tau) \vert \hat{H} \vert \Psi(\tau) \rangle 
\nonumber \\
        & = & \int^{\infty}_{-\infty} \left\{ 
\psi^{\ast}_1 \hat{h}_{1\zeta} \psi_1 + \psi^{\ast}_2 \hat{h}_{2\zeta} \psi_2 
\right\} d\zeta   
\nonumber \\
        &   & - j \int^{\infty}_{-\infty} \left\{ 
\psi^{\ast}_1 \psi_2 + \psi^{\ast}_2 \psi_1 \right\} d\zeta   \: .
\label{drivpol:eq:twentyfivefour}
\end{eqnarray}
It can be shown by standard means that $E(\tau)$ is real for all times $\tau$.

\section{\label{drivpol:sec:four}
Limit cases
}

In this section, some limit cases of general model~(\ref{drivpol:eq:eleven}) are reviewed. Consider first zero carrier-ion interaction, $\lambda = 0$. That splits the system in two independent subsystems: (i) a free ion interacting with an external field, and (ii) a free carrier hopping between two sites and interacting with an external field. Solution to Eq.~(\ref{drivpol:eq:eleven}) is sought in a factorized form: $\psi_{1,2}(\zeta,\tau) = \phi(\zeta,\tau) \cdot \chi_{1,2}(\tau)$. Equation~(\ref{drivpol:eq:eleven}) decouples into one equation for $\phi$:
\begin{equation}
i \frac{\partial \phi(\zeta,\tau)}{\partial \tau} = 
\left\{ - \frac{1}{2} \frac{\partial^2}{\partial \zeta^2} + \frac{1}{2} \, \zeta^2 
- z_i \alpha_i \zeta \sin{(\omega \tau)} \right\} \phi(\zeta,\tau) \: , 
\label{drivpol:eq:twentysix}
\end{equation}
and two coupled equations for $\chi_{1,2}$
\begin{eqnarray}
i \frac{\partial \chi_1}{\partial \tau} & = &   
  z_c \alpha_c \sin{(\omega\tau)} \, \chi_1 - j \, \chi_2 \: , 
\label{drivpol:eq:twentyseven} \\
i \frac{\partial \chi_2}{\partial \tau} & = & 
- j \, \chi_1 - z_c \alpha_c \sin{(\omega\tau)} \, \chi_2 \: .
\label{drivpol:eq:twentyeight}
\end{eqnarray}
Equation~(\ref{drivpol:eq:twentysix}) admits an explicit solution by means of Husimi substitution~\cite{Husimi1953,Kerner1958} that maps a forced harmonic oscillator onto a free oscillator. For example, the lowest ($m = 0$) Floquet state has the form
\begin{equation}
\phi_0(\zeta,\tau) = e^{-i \varepsilon_0 \tau + i S(\zeta,\tau)} 
\frac{1}{\pi^{1/4}} \, e^{-\frac{1}{2}[\zeta - \eta(\tau)]^2} \: , 
\label{drivpol:eq:twentynine}
\end{equation}
with
\begin{equation}
\varepsilon_0 = \frac{1}{2} - \frac{ (z_i \alpha_i)^2 }{4 ( 1 - \omega^2 )}  \: , 
\label{drivpol:eq:thirty}
\end{equation}
\begin{equation}
S(\zeta,\tau) = \frac{z_i \alpha_i \, \omega}{ 1 - \omega^2 } \, \zeta \cos{(\omega\tau)}
- \frac{(z_i \alpha_i)^2 ( 1 + \omega^2 ) }{ 8 \omega ( 1 - \omega^2 )^2 } \sin{( 2 \omega \tau )} \: , 
\label{drivpol:eq:thirtyone}
\end{equation}
\begin{equation}
\eta(\tau) = \frac{ z_i \alpha_i \, \omega }{ 1 - \omega^2 } \, \sin{(\omega\tau)} \: . 
\label{drivpol:eq:thirtytwo}
\end{equation}
This explicit formula can be used to validate the numerical method described in the preceding section. In case of resonant excitation, $\omega = 1$, no stable periodic solution is possible. Instead, the amplitude of wave function oscillations $\eta(\tau)$ grows linearly with time. If the oscillator is in its ground state when the field is turned on, the time-dependent solution still has the form of Eq.~(\ref{drivpol:eq:twentynine}) but with $\varepsilon_0 = \frac{1}{2}$ and new functions $\eta(\tau)$ and $S(\tau)$: 
\begin{equation}
\eta_{\rm res}(\tau) = \frac{ z_i \alpha_i }{ 2 } \, \left( \sin{\tau} - \tau \cos{\tau} \right) , 
\label{drivpol:eq:thirtytwoone}
\end{equation}
\begin{eqnarray}
& & S_{\rm res}(\zeta,\tau) = \frac{z_i \alpha_i }{ 2 } \, \zeta \, \tau \sin{\tau} 
\nonumber \\
& & + \frac{(z_i \alpha_i)^2 }{ 16 }  
\left( \tau - \frac{3}{2} \sin{2\tau} + \tau^2 \sin{2\tau} + 2\tau \cos{2\tau} \right) . 
\label{drivpol:eq:thirtyonetwo}
\end{eqnarray}
\begin{figure}[t]
\includegraphics[width=0.48\textwidth]{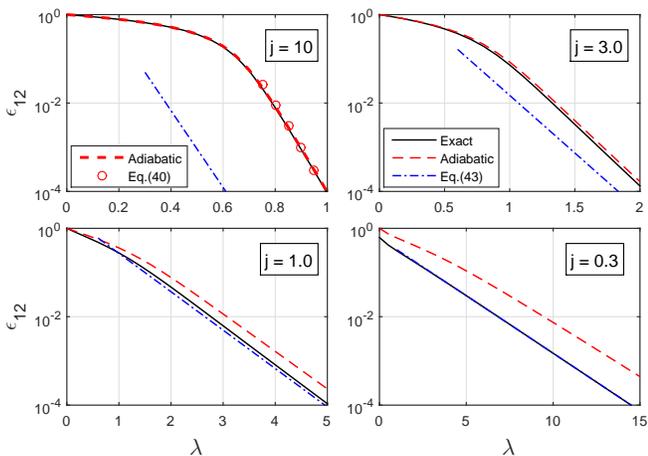}
\caption{(Color online) Energy split of the two lowest stationary eigenstates of the two-site polaron for several adiabaticity parameters $j$. Exact numeric eigenvalues from Eq.~(\ref{drivpol:eq:eleven}) (for $\alpha_i = \alpha_e = 0$) and adiabatic numeric eigenvalues from Eq.~(\ref{drivpol:eq:thirtythree}) are shown by solid and dashed lines, respectively. The exact and adiabatic eigenstates are nearly indistinguishable at $j = 10$. The same is true for the exact and anti-adiabatic solution, Eq.~(\ref{drivpol:eq:thirtysix}) shown by dot-dashed lines, at $j = 0.3$. In all cases, exact eigenvalues lie between the adiabatic and anti-adiabatic limits. Notice also good match between numeric adiabatic values and the instanton formula, Eq.~(\ref{drbip:eq:thirteen}), for large $j$ and $\lambda$.}
\label{drivpol:fig:three}
\end{figure}

Equations~(\ref{drivpol:eq:twentyseven}) and (\ref{drivpol:eq:twentyeight}) and their analogues in quantum optics have been studied in several papers~\cite{Bloch1940,Shirley1965,Aravind1984,Agarwal1994,Raghavan1996} but no analytical solution has been reported. (Somewhat surprisingly, the more general problem of an infinite chain in a periodically varying electric field {\em is} exactly solvable, providing analytical description of dynamic localization.~\cite{Dunlap1986}) To gain some insight, we show several numerical solutions in Fig.~\ref{drivpol:fig:two}. At low frequencies, $\omega \ll j$, the on-site energies change slowly compared to inter-site hopping. When the levels are out of resonance, the carrier gets confined to one of the sites and the frequency of oscillations increases, as can be seen in the left panels of Fig.~\ref{drivpol:fig:two}. At high frequencies, $\omega \gg j$, fast oscillations of on-site energies average to zero and the dynamics approaches that of a free carrier with an intersite tunneling amplitude $j$, see Fig.~\ref{drivpol:fig:two} on the right. At intermediate frequencies, carrier dynamics is complex, as illustrated in Fig.~\ref{drivpol:fig:two}, center. 
   
In the absence of an external field, $\alpha_i = \alpha_c = 0$, the general problem, Eq.~(\ref{drivpol:eq:eleven}), reduces to a free two-site polaron. Note that although the present model involves an intersite ion that couples to both sites simultaneously, the model can be mapped to the two-site Holstein polaron that has been studied in detail.~\cite{Ranninger1992,Kabanov1994,Holstein1959b} A relevant quantity is the energy split of the lowest level pair, as it defines polaron ``mass'' and tunneling frequency in the undriven case [$f_{\rm tun} = \triangle \epsilon_{12}/(2\pi)$]. In this paper, the ground state energy and the first excited state energy are computed by discretizing polaron Hamiltonian of Eq.~(\ref{drivpol:eq:eleven}) in $\zeta$ space and directly diagonalizing a resulting matrix, see Fig.~\ref{drivpol:fig:three}. Tunneling frequencies thus obtained are later used as a reference for the driven case. 

There are two well understood limits of the (undriven) two-site polaron. In the adiabatic limit,~\cite{Holstein1959b,Emin1976,Kabanov1993,Kabanov1994} $j \gg 1$, the two-function Schr\"odinger equation, Eq.~(\ref{drivpol:eq:eleven}), is reduced to a one-function equation with Hamiltonian
\begin{equation}
\hat{h}_{\rm ad} =  
- \frac{1}{2} \frac{\partial^2}{\partial \zeta^2} + w(\zeta)  \: ,
\label{drivpol:eq:thirtythree}
\end{equation}
where $w(\zeta)$ is the adiabatic potential, Eq.~(\ref{drivpol:eq:nineteen}). Eigenvalues of $\hat{h}_{\rm ad}$ can be found numerically by discretizing Eq.~(\ref{drivpol:eq:thirtythree}) using standard rules. Level splitting can also be estimated with the instanton technique,~\cite{Coleman1977,Rajaraman1982,Kleinert2004} leading to the following result:~\cite{Kornilovitch2016,Kabanov1994} 
\begin{equation}
\triangle \epsilon^{\rm inst}_{12} = 
\sqrt{\frac{8 j}{\pi \lambda}} \, F_1 \left( \frac{\lambda}{\lambda_{\rm cr}} \right) \,
e^{- \frac{j}{2\lambda} F_2 \left( \frac{\lambda}{\lambda_{\rm cr}} \right) } \: , 
\label{drbip:eq:thirteen}
\end{equation}
\begin{equation}
F_1(x) =  
\frac{ x^2 \left( 1 - x^{-2} \right)^{5/4} }
{\left[ x \left( 1 + \sqrt{1 - x^{-2}} \right) \right]^{ \sqrt{1 - x^{-2}} } } \: , 
\label{drbip:eq:fourteen}
\end{equation}
\begin{equation}
F_2(x) = x^2 \sqrt{1 - x^{-2}} - \log{ \left[ x \left( 1 + \sqrt{1 - x^{-2}} \right) \right] } \: ,
\label{drbip:eq:fifteen}
\end{equation}
where $\lambda_{\rm cr} = \frac{1}{2}$. Figure~\ref{drivpol:fig:three} compares Eq.~(\ref{drbip:eq:thirteen}) with numerical diagonalization of adiabatic Hamiltonian $\hat{h}_{\rm ad}$ and of full Hamiltonian, Eq.~(\ref{drivpol:eq:eleven}). 

In the opposite, anti-adiabatic limit, $j \ll 1$, a fast ion instantaneously follows carrier transitions. Level splitting is derived from the sudden approximation~\cite{Tjablikov1951,Sewell1958,Holstein1959a} 
\begin{equation}
\triangle \epsilon_{12} = 2 j \, e^{ - 2 j \lambda }  \: .
\label{drivpol:eq:thirtysix}
\end{equation}

We now turn to the full time-dependent problem, Eq.~(\ref{drivpol:eq:eleven}), with nonzero $\alpha_i$ and $\alpha_c$. We begin with the adiabatic limit $j \gg 1$.

\begin{figure}[t]
\includegraphics[width=0.48\textwidth]{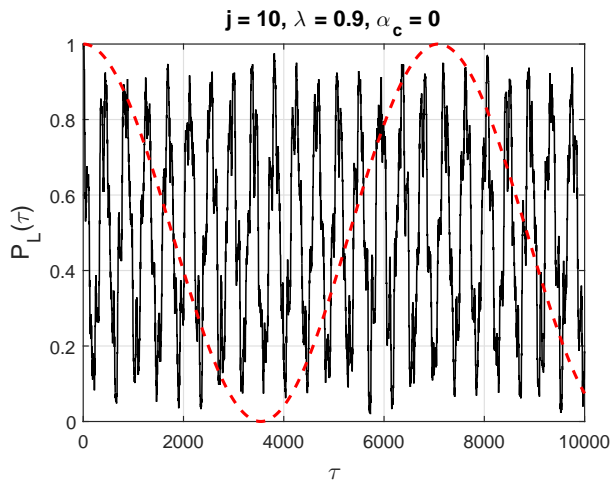}
\caption{(Color online) Time evolution of two-site polaron for $j = 10$, $\lambda = 0.9$ and $\alpha_c = 0.0$. The dashed line is undriven case $\alpha_i = 0$, while the solid line is driven case at $\alpha_i = 0.05$ and $\omega = 0.71$.}
\label{drivpol:fig:four}
\end{figure}

\section{\label{drivpol:sec:five}
Two-site driven polaron: Adiabatic case
}

Effects of direct ion-field interaction, $\alpha_i \neq 0$, $\alpha_c = 0$, are considered first. Figure~\ref{drivpol:fig:four} compares time evolution of the driven and undriven polaron for $j = 10$ and $\lambda = 0.9$. Period of undriven oscillations is $T_0 = 7080$ which is consistent with the lowest doublet split of $\triangle \epsilon_{12} = 8.81 \cdot 10^{-4}$ [formula (\ref{drbip:eq:thirteen}) is accurate to 5.7\%]. In contrast, the driven case shows much faster oscillations with a period $T \approx 410$, i.e. a decrease of 17.3 times. Notice that the driven $P_L(\tau)$ is not a clean sinusoid but rather a complex function with multiple overtones. 

\begin{figure}[t]
\includegraphics[width=0.48\textwidth]{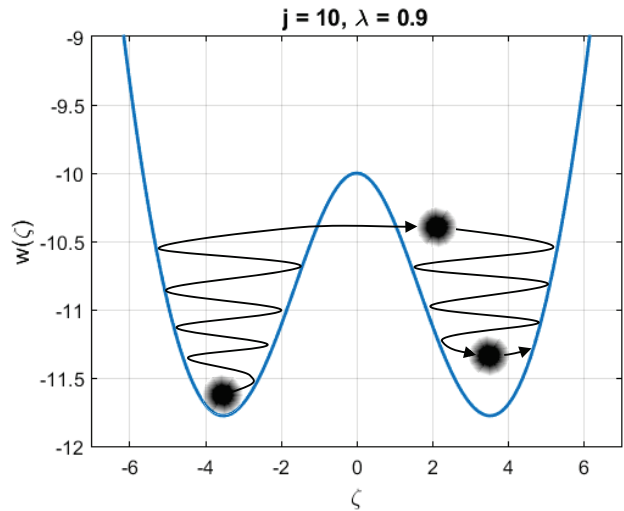}
\caption{(Color online) Proposed mechanism of field-enhanced polaron tunneling. The double-well function is adiabatic potential, Eq.~(\ref{drivpol:eq:nineteen}), for $j = 10$ and $\lambda = 0.9$. The two minima are located at $\zeta_0 = \pm 3.52$. Dots with fuzzy edges symbolize an ion wave function at different times. Starting from the bottom of the left well, the ion absorbs energy from the field and rises to the top of potential barrier. After tunneling to the right well, the ion loses energy and slows down. Then the cycle repeats.}
\label{drivpol:fig:five}
\end{figure}

A possible mechanism behind faster oscillations is now discussed. The argument is based on the property of quantum oscillators that the centroid of a driven wave function follows a classical equation of motion.~\cite{Husimi1953} Consider an ion near the bottom of the adiabatic potential, as shown in Fig.~\ref{drivpol:fig:five}. Frequency of small-amplitude oscillations is $\tilde{\omega} = \sqrt{ 1 - (\lambda_{\rm cr}/\lambda)^2 }$, see Eq.~(\ref{drivpol:eq:twentyone}). The classical equation of motion at resonance reads 
\begin{equation}
\ddot{x} + \tilde{\omega}^2 \, x = z_i \alpha_i \sin{(\tilde{\omega} \tau)}  \: . 
\label{drivpol:eq:fortyone}
\end{equation}
An explicit solution with arbitrary starting point $x_0$ and zero initial velocity is  
\begin{equation}
x(\tau) = x_0 \cos{(\tilde{\omega} \tau)} + \frac{z_i \alpha_i}{2 \tilde{\omega}^2} 
\left[ \, \sin{(\tilde{\omega} \tau)} - (\tilde{\omega} \tau) \cos{(\tilde{\omega} \tau)} \, \right] \: .
\label{drivpol:eq:fortytwo}
\end{equation}
The oscillation-averaged total energy is
\begin{equation}
\langle \varepsilon(\tau) \rangle = 
\frac{1}{8} \left( z_i \alpha_i \tau - 2 \tilde{\omega} \, x_0 \right)^2
+ \frac{ ( z_i \alpha_i )^2 }{8 \, \tilde{\omega}^2}  \: .
\label{drivpol:eq:fortythree}
\end{equation}
Long-term behavior of $\langle \varepsilon(\tau) \rangle$ is quadratic growth. However, short-term details depend on starting position $x_0$. For positive $x_0$, energy initially decreases and oscillations stop before picking up again. On the basis of this observation, the following mechanism can be proposed, see Fig.~\ref{drivpol:fig:five}. Initially, the ion is near the bottom of the left well of the adiabatic potential. The external field drives the ion in resonance and its energy rises. As energy approaches the top of the potential barrier, quantum-mechanical tunneling probability increases exponentially and eventually the ion tunnels under the barrier into the right well. There, the ion finds itself out-of-phase with the field (the in-phase position would be on the {\em right} side of the right well) and starts losing energy. Once the ion descends to the bottom of the right well, the process repeats but in the opposite direction.

Reality is more complicated. One complication comes from non-harmonicity of the adiabatic potential. As can be inferred from Eq.~(\ref{drivpol:eq:nineteen}), the instantaneous frequency {\em decreases} with the amplitude of oscillations. As a result, an external field cannot always be in resonance. One can only speak of an ``effective'' or ``average'' resonance for the duration of the process. It is clear that the rate of energy absorption or loss is less than for ideal resonance. Another complication is the gradual nature of quantum-mechanical tunneling. Numerical analysis shows that under-barrier tunneling is not a sharp process. Rather, the full wave function is nonzero in both wells at all times, so the ``start'' or ``end'' of tunneling is not easy to identify.      

\begin{figure}[t]
\includegraphics[width=0.48\textwidth]{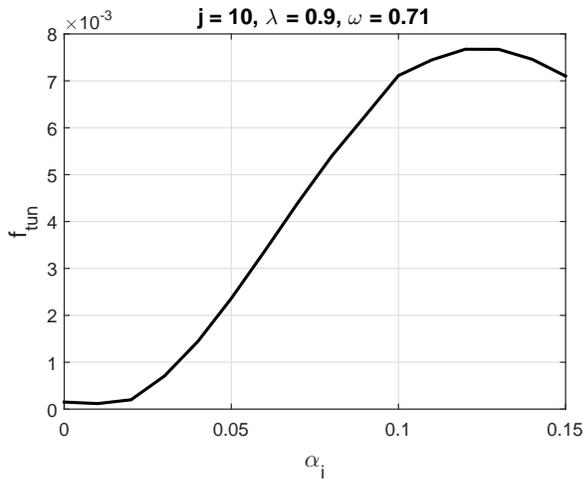}
\caption{ Tunneling frequency $f_{\rm tun}$ versus ion-field coupling $\alpha_i$ for $j = 10$ and $\lambda = 0.9$. $f_{\rm tun}$ has been determined from the maximum of Fourier power spectrum of $P_L(\tau)$. }
\label{drivpol:fig:seven}
\end{figure}

The qualitative argument presented above leads to an important conclusion. According to Eq.~(\ref{drivpol:eq:fortytwo}), the rate of amplitude increase is proportional to coupling constant $\alpha_i$. If one assumes that tunneling takes place when energy reaches a certain value, and by association when the amplitude reaches a certain value, then the tunneling condition will be reached faster at larger $\alpha_i$. This implies that the frequency of {\em driven} tunneling is directly proportional to the field strength and to the square root of laser intensity
\begin{equation}
f_{\rm tun} \propto \alpha_i \propto \sqrt{I} \: .
\label{drivpol:eq:fortyfour}
\end{equation}
To verify this prediction, we show $f_{\rm tun}(\alpha_i)$ computed for $j = 10$ and $\lambda = 0.9$ in Fig.~\ref{drivpol:fig:seven}. $f_{\rm tun}$ has been determined from the location of the largest peak in a Fourier power spectrum of $P_{L}(\tau)$. Relation (\ref{drivpol:eq:fortyfour}) approximately holds at  intermediate $0.03 < \alpha_i < 0.10$. There is a local minimum at weak couplings, $\alpha_i \approx 0.01$, which implies that a very weak coupling initially disrupts quantum-mechanical tunneling and decreases the tunneling rate. In the opposite limit of strong coupling, $\alpha_i > 0.1$, oscillations become highly irregular without a dominant harmonic in the Fourier transform.

Notice that experimentally achievable field strengths,~\cite{Hu2014} $\alpha_i = 0.02$, lie near the lower end of the linearity region, according to the present calculation. This suggests that a further increase in the field strength should systematically raise $f_{\rm tun}$. This effect may have implications for superconductivity in the cuprates, as was suggested in Ref.~[\onlinecite{Kornilovitch2016}].   

\begin{figure}[t]
\includegraphics[width=0.48\textwidth]{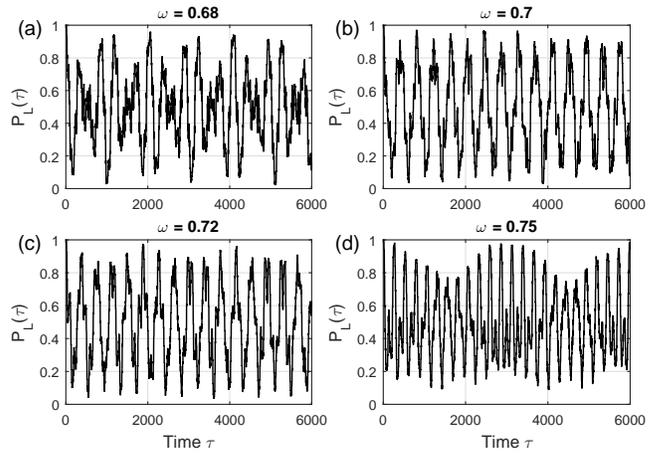}
\caption{Time evolution of two-site polaron for several driving frequencies. Model parameters are $j = 10$, $\lambda = 0.9$, $\alpha_i = 0.05$, and $\alpha_c = 0.0$. The ``optimal'' driving frequency for this case is $\omega = 0.71$, for which data were shown in Fig.~\ref{drivpol:fig:four}. }
\label{drivpol:fig:sevenone}
\end{figure}

In real crystals, local ion frequencies may be shifted by dispersion, lattice imperfections, temperature, and other factors. Therefore it is important to understand the sensitivity of $f_{\rm tun}$ increase to variations of ion frequency $\Omega$. In the present model, $\Omega^{-1}$ is taken to be a unit of time, so one should consider sensitivity to external driving frequency $\omega$ instead. Figure~\ref{drivpol:fig:sevenone} presents $P_L(\tau)$ for several values of $\omega$. One can observe that enhanced polaron tunneling persists for detunings of up to several percent of $\Omega$ on both sides of the ``optimal'' frequency $\omega = 0.71$. (Which of course renders the choice of optimal frequency itself somewhat arbitrary.) For larger detunings, $P_L(\tau)$ assumes more complex beat-like shapes which makes it difficult to interpret the motion as periodic tunneling between two potential wells. An important conclusion, however, is that polaron delocalization is a robust effect that exists in a finite interval of ion and driving frequencies, and as such should be observable in real materials.

\begin{figure}[b]
\includegraphics[width=0.48\textwidth]{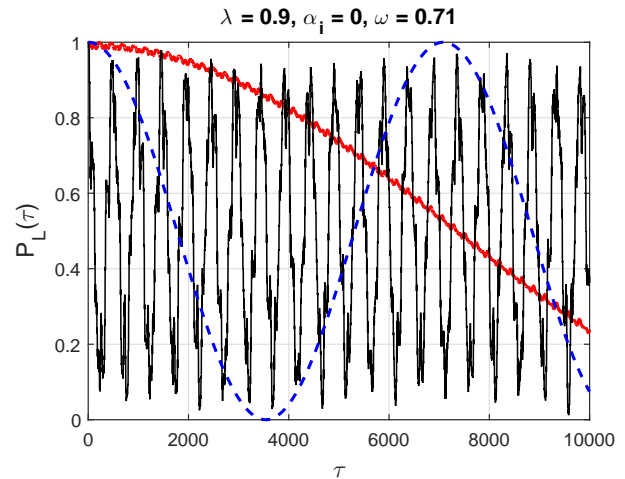}
\caption{(Color online) Time evolution of two-site polaron with carrier-field drive for $j = 10$, $\lambda = 0.9$, $\omega = 0.71$. There is no ion-field coupling, $\alpha_i = 0$. Dashed line is the undriven case, $\alpha_e = 0.0$, thick slow-varying line is $\alpha_e = 0.2$, and thin fast-varying line is $\alpha_e = 0.8$. }
\label{drivpol:fig:eight}
\end{figure}

So far, only interaction between the external field and ion has been taken into account. Now we consider the opposite case of carrier-field interaction in the {\em absence} of ion-field interaction: $\alpha_e \neq 0$ and $\alpha_i = 0$. Figure~\ref{drivpol:fig:eight} compares $P_{L}(\tau)$ for several $\alpha_e$. The overall $f_{\rm tun}(\alpha_e)$ dependence is similar to the $f_{\rm tun}(\alpha_i)$ behavior: initially, at small $\alpha_e$, the frequency decreases relative to the undriven case before increasing at larger $\alpha_e$.

\begin{figure}[t]
\includegraphics[width=0.42\textwidth]{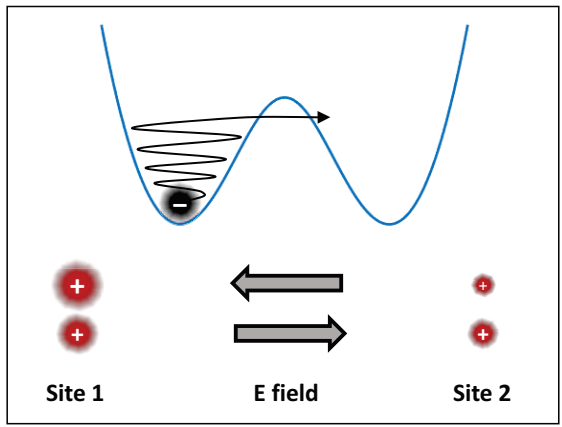}
\caption{ (Color online) Mechanism of enhanced polaron tunneling for $\alpha_e \neq 0$. Oscillating electric field changes partial carrier weights on the two sites. The weights are indicated by positively charged dots. By virtue of carrier-ion coupling, oscillating carrier density excites the negatively charged ion at the bottom of the potential well.}
\label{drivpol:fig:nine}
\end{figure}

A possible underlying mechanism is now discussed, see Fig.~\ref{drivpol:fig:nine}. In the adiabatic limit, $j \gg 1$, $ j \gg \omega$, the carrier wave function equilibrates between the two sites for any instantaneous position of the ion and instantaneous value of the field. The time-dependent field ``rocks'' the carrier density between the two sites. In turn, the oscillating carrier density pulls the ion with variable strength and excites the latter in its own well if the field frequency is close to the ion frequency. Thus instead of direct excitation by the field, the ion is driven into resonance by oscillating carrier density and nonzero carrier-ion interaction.

Fluctuations of carrier density occur on a background set by the much larger bare kinetic energy $J$. As a result, influence of the external field is effectively reduced. In order to estimate this effect, solve the two-site problem for small $\alpha_e \ll j$. In the first order, one obtains for the wave functions
\begin{equation}
\frac{ \psi_2(\tau) }{ \psi_1(\tau) } = 
\left( 2 + \frac{z_e \alpha_e \sin{\omega\tau}}{\lambda j} \right)
\left( \lambda - \sqrt{ \lambda^2 - \lambda^2_{\rm cr} } \right)  \: ,
\label{drivpol:eq:fortysix}
\end{equation}
and for the densities
\begin{eqnarray}
\vert \psi_1(\tau) \vert^2 & = &
\frac{1}{8\lambda \left( \lambda - \sqrt{ \lambda^2 - \lambda^2_{\rm cr} } \right)} 
- \triangle n(\tau) \: ,
\label{drivpol:eq:fortyseven} \\
\vert \psi_2(\tau) \vert^2 & = &
\frac{ \left( \lambda - \sqrt{ \lambda^2 - \lambda^2_{\rm cr} } \right)}{2\lambda} 
+ \triangle n(\tau) \: ,
\label{drivpol:eq:fortyeight}
\end{eqnarray}
where 
\begin{equation}
\triangle n(\tau) = \frac{z_e \alpha_e \sin{\omega\tau}}{ 16 j \lambda^3 } \: . 
\label{drivpol:eq:fortynine}
\end{equation}
(Note that $\vert \psi_1(\tau) \vert^2 + \vert \psi_2(\tau) \vert^2 = 1$.) Thus, a carrier-field interaction plus a carrier-ion interaction is equivalent to an ion-field interaction with some effective coupling constant $\tilde{\alpha}_{i}$. Recasting carrier-ion interaction, Eq.~(\ref{drivpol:eq:four}), with $\triangle n$ from Eq.~(\ref{drivpol:eq:fortynine}) into ion-field interaction form, Eq.~(\ref{drivpol:eq:six}), one obtains
\begin{equation}
\tilde{\alpha}_i = \frac{z_e}{z_i} \frac{\alpha_e}{ \sqrt{ 32 j \, \lambda^5 } } \: . 
\label{drivpol:eq:fifty}
\end{equation}
For $j = 10$ and $z_i = - 2$, the last formula changes from $\tilde{\alpha}_i \approx - \frac{\alpha_e}{6}$ at $\lambda = \frac{1}{2}$ to $\tilde{\alpha}_i \approx - \frac{\alpha_e}{36}$ at $\lambda = 1$. Using the estimates given in Sec.~\ref{drivpol:sec:two}, $\alpha_e = 1.5$ and $\alpha_i = 0.02$, one concludes that $\tilde{\alpha}_i > \alpha_i$, that is the indirect light-ion interaction is at least comparable to the direct interaction and may even be dominant. One should note that precise values of $\alpha_e$ and $\alpha_i$ depend on the local crystal structure and fine details of the dielectric response of a particular solid. These topics are beyond the scope of the present work and for this reason $\alpha_e$ and $\alpha_i$ will continue to be treated here as phenomenological parameters.      

\begin{figure}[t]
\includegraphics[width=0.48\textwidth]{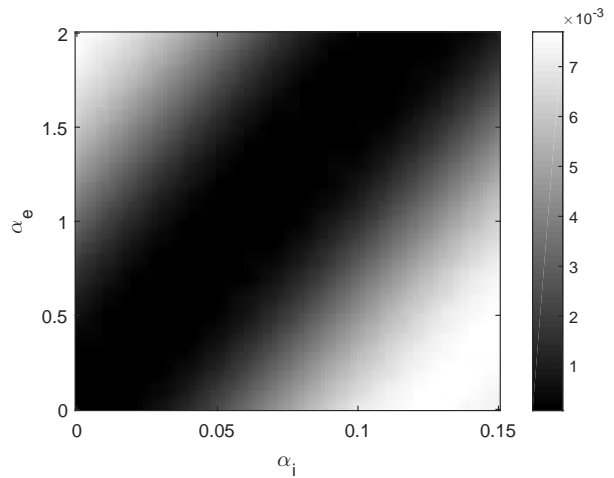}
\caption{Polaron tunneling frequency $f_{\rm tun} = 1/T$ vs. two coupling constants $\alpha_i$ and $\alpha_e$ for $j = 10$, $\lambda = 0.9$ and $\omega = 0.71$. Note a broad depression along the line $\alpha_e = (22-25) \alpha_i$ which is in agreement with Eq.~(\ref{drivpol:eq:fifty}). To create this figure, $f_{\rm tun}$ was first computed on a $(16 \times 21)$ mesh and then interpolated to a $(121 \times 161)$ mesh. }
\label{drivpol:fig:ten}
\end{figure}

Nonetheless, one general conclusion can be reached by examining Fig.~\ref{drivpol:fig:nine}: the direct and indirect ion-field interactions are {\em always} of opposite signs. For example, in case of attraction (depicted in the figure), the same field pulls the ion in one direction while modifying carrier density to pull the ion in the opposite direction. This can also be seen in Eq.~(\ref{drivpol:eq:fifty}) that becomes negative whenever $z_e$ and $z_i$ are of opposite signs. The main conclusion remains true in case of carrier-ion repulsion.  

Thus the two interactions always compete and the combined effect is always smaller than either of the two. This competing nature is confirmed by numerical calculations summarized in Fig.~\ref{drivpol:fig:ten}. Shown is the polaron tunneling frequency, as determined from the location of the largest peak in a power spectrum of $P_L(\tau)$, as a function of $\alpha_i$ and $\alpha_e$. It is evident from the figure that $f_{\rm tun}$ is in fact a function of $\alpha^{\prime}_i = \alpha_i - c^{-1} \alpha_e$, where $c \approx 22 - 25$ is a numerical coefficient that is in agreement with Eq.~(\ref{drivpol:eq:fifty}). Given that neither $\alpha_i$ nor $\alpha_e$ is known (unless detailed first-principle calculations of the dielectric response are performed) it is convenient to lump both interactions in one effective ion-field interaction with a new phenomenological coupling constant $\alpha^{\prime}_i$. Therefore, $\alpha_i = \alpha^{\prime}_i$ and $\alpha_e = 0$ will be assumed for the rest of the paper. Again, we choose to deal with an effective ion-field interaction rather than carrier-field interaction because of easier visualization.

\begin{figure}[t]
\includegraphics[width=0.48\textwidth]{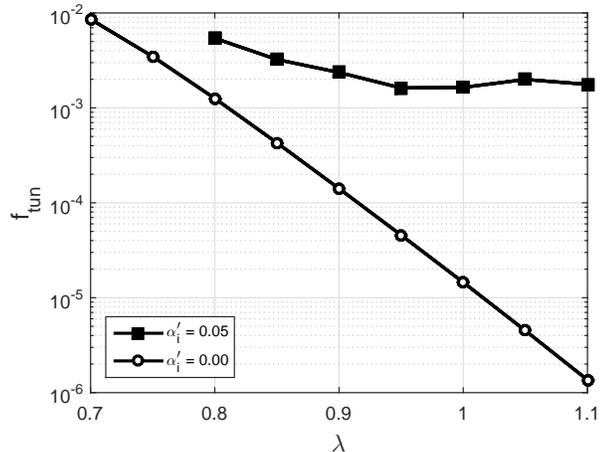}
\caption{Polaron tunneling frequency $f_{\rm tun} = 1/T$ for $j = 10$. In the undriven case, $f_{\rm tun}$ declines exponentially with $\lambda$, while it is approximately constant in the $\alpha^{\prime}_i = 0.05$ driven case.}
\label{drivpol:fig:eleven}
\end{figure}

We now return to the main subject of this paper: {\em exponential} increase of the tunneling frequency under a driving field. Figure~\ref{drivpol:fig:eleven} compares $f_{\rm tun}$ of the driven and undriven polarons as a function of $\lambda$. The undriven line is essentially the same that was shown earlier in Fig.~\ref{drivpol:fig:three} but now computed not from the spectrum of a stationary Schr\"odinger equation but from Fourier analysis of the time-dependent Schr\"odinger equation, and divided by $2\pi$. The two methods produce numbers that match within less than 1\%. The undriven case shows the familiar exponential decrease with $\lambda$ reflecting a deepening potential barrier between two minima. In contrast, $f_{\rm tun}$ of the driven case remains approximately constant. This is understandable, since a small deepening of the potential results in a proportionally small increase of the time the ion needs to gain enough energy to get over the barrier. In other words, in the driven case, $f_{\rm tun}$ is more a function of the field strength rather than details of the potential.        

An important consequence is that the ratio between driven and undriven $f_{\rm tun}$ grows {\em exponentially} in the deep polaron regime. In the example of Fig.~\ref{drivpol:fig:eleven}, the driven frequency exceeds the undriven one by two orders of magnitude at $\lambda = 1.0$ and by three orders at $\lambda = 1.1$. Thus field assistance promotes polaron tunneling and increases polaron mobility. This effect has important consequences for transport properties of polaron systems.

\section{\label{drivpol:sec:six}
Non-adiabatic cases 
}

\subsection{\label{drivpol:sec:sixone}
Slow carriers, fast ions: $j \ll 1$ 
}

\begin{figure}[t]
\includegraphics[width=0.48\textwidth]{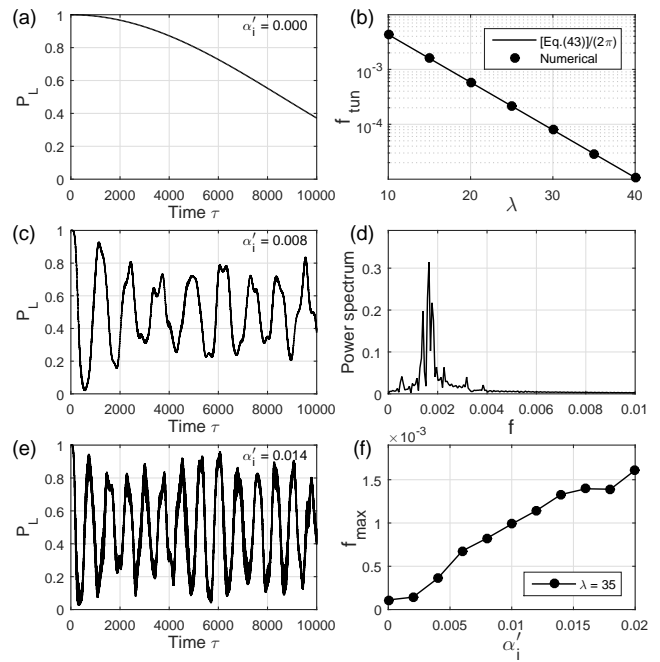}
\caption{Driven polaron in the anti-adiabatic regime, $j = 0.1$, $\lambda = 35$ and $\omega = 1.0$. (a) Left side probability $P_L(\tau)$ in the undriven case. (b) Tunneling frequency $f_{\rm tun} = 1/T$ of the undriven polaron compared with the energy split, Eq.~(\ref{drivpol:eq:thirtysix}), divided by $(2\pi)$. (c) $P_L(\tau)$ in the driven case, for $\alpha^{\prime}_i = 0.008$. (d) Fourier spectrum of the time series shown in panel (c). (e) $P_L(\tau)$ for a larger coupling constant $\alpha^{\prime}_i = 0.014$, showing faster oscillations. (f) Tunneling frequency vs. $\alpha^{\prime}_i$ showing an approximately linear dependence. }
\label{drivpol:fig:twelve}
\end{figure}

We now turn to the opposite anti-adiabatic case $j \ll 1$. In this regime, ion oscillations are fast and carrier transitions between the sites are rare. When the carrier is confined to one of the two sites, the ion shifts to a new equilibrium position toward the carrier by $( 2 j \lambda)^{1/2}$. When the carrier tunnels to the second site, the displacement must reverse. Overlap of ion wave functions separated by $2 ( 2 j \lambda)^{1/2}$ leads to an exponentially small level splitting given by Eq.~(\ref{drivpol:eq:thirtysix}). Numerical solution confirms this physical picture. Figure~\ref{drivpol:fig:twelve}(a) shows temporal evolution of an undriven polaron for $j = 0.1$ and $\lambda = 35$. The left-space probability has an ideal sine-wave shape as expected for a closely split level dublet. The numerical time period is in perfect agreement with the analytical formula, as shown in Fig.~\ref{drivpol:fig:twelve}(b).    

An external electric field drives the ion in resonance. As the oscillation amplitude rises in accordance with Eq.~(\ref{drivpol:eq:fortytwo}), the overlap integral grows allowing for more frequent transitions. Increased tunneling rate between the two states is evident in Fig.~\ref{drivpol:fig:twelve}(c). Figure~\ref{drivpol:fig:twelve}(d) of the same figure shows the corresponding Fourier spectrum with a prominent peak near 0.00165. Different from the adiabatic case, in the $j \ll 1$ limit the ionic potential remains undistorted and harmonic. As a result, the most effective driving frequency is simply $\omega = 1.0$. This is confirmed by numerical results.         

Similarly to the adiabatic case, stronger external fields cause more frequent tunneling. This can be seen by comparing the time series in Figs.~\ref{drivpol:fig:twelve}(c) and \ref{drivpol:fig:twelve}(e). In the latter case, the field is almost twice as strong (0.014 vs. 0.008), which leads to faster oscillations. The overall $f_{\rm tun}(\alpha^{\prime}_i)$ dependence is roughly linear, as shown in Fig.~\ref{drivpol:fig:twelve}(f). The physical reason is the same as before: the amount of time needed to reach an amplitude at which tunneling takes place ``comfortably'' is inversely proportional to $\alpha^{\prime}_i$ and ${\cal E}$.

\begin{figure}[t]
\includegraphics[width=0.48\textwidth]{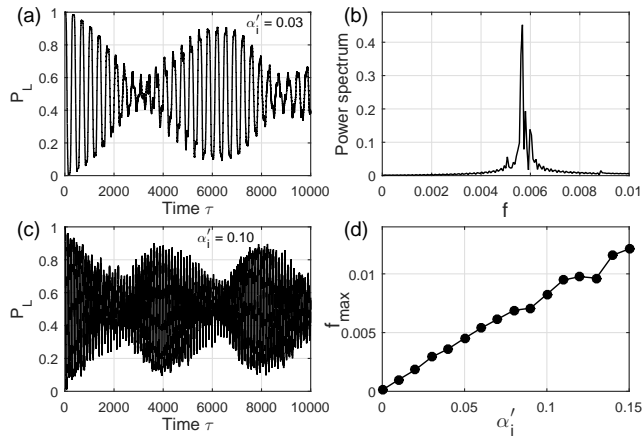}
\caption{Driven polaron at intermediate phonon frequencies, $j = 1.0$, $\lambda = 5.0$, and $\omega = 0.99$. (a) $P_L(\tau)$ for $\alpha^{\prime}_i = 0.03$. (b) Fourier spectrum of $P_L(\tau)$ shown in (a). (c) $P_L(\tau)$ for a larger coupling constant $\alpha^{\prime}_i = 0.10$, showing faster oscillations. (d) Tunneling frequency vs. $\alpha^{\prime}_i$ showing an approximately linear dependence. }
\label{drivpol:fig:thirteen}
\end{figure}

\subsection{\label{drivpol:sec:sixtwo}
Intermediate frequencies: $j \sim 1$ 
}

As a typical example of intermediate phonon frequencies we consider $j = 1.0$. The polaron is expected to behave between the $j \gg 1$ and $j \ll 1$ limits. In particular, an optimal driving frequency should be less than 1.0 but not as much as in the adiabatic regime. Figure~\ref{drivpol:fig:thirteen} shows a sample of numerical results obtained for $\lambda = 5.0$. A scan over driving frequencies found that the most prominent tunneling occurs at $\omega = 0.99$, i.e., this case is closer to the anti-adiabatic limit. The figure reveals the now familiar physical picture. The undriven case (not shown) corresponds to a deep polaron regime with a large tunneling period of $T \approx 85\,880$. An external force promotes tunneling and increases $f_{\rm tun}$ by 2-to-3 orders of magnitude. $f_{\rm tun}$ increases linearly with the field's strength in accordance with the physical mechanism described in previous sections.

\section{\label{drivpol:sec:seven}
Summary and discussion
}

The ability to drive ions in resonance by external laser fields is a powerful technique of experimental solid state physics. In systems where ions are strongly coupled to charge carriers or other degrees of freedom, direct excitation of ions must lead to large and measurable changes in other subsystems. In this paper, we investigated one model system where such effects are particularly strong: the small lattice polaron. By definition, a lattice polaron is a charge carrier that interacts with surrounding ions so strongly that it deforms the lattice and displaces the ions from their equilibrium positions by finite amounts. If those ions are agitated by a different source, in our case by an external laser field, that should change the balance of forces between the ions and the carrier. Specifically, the energy barrier that prevents the carrier from tunneling between lattice sites will be affected. That should lead to {\em exponentially} large changes in polaron tunneling rates, which may be detectable through transport or optical response. 

In the absence of analytical solutions even for the simplest cases, we employed direct step-by-step integration of the time-dependent Schr\"odinger equation as the main analysis tool. A two-site polaron model has been chosen as being sufficient to reveal the essential physics. Our findings can be summarized as follows.  

(i) Increase of polaron tunneling rate has been found at all ion frequencies $\Omega$ (as compared to the bare transfer integral $J$), as long as the field is appropriately tuned to the ion resonance. The effect exists in a finite range of driving frequencies. In accordance with general polaron theory, an optimal driving frequency has been found to be $\omega = \Omega$ in the anti-adiabatic limit $J \ll \Omega$, and gradually decreasing to $\omega < \Omega$ in the adiabatic limit $J \gg \Omega$. 

(ii) Resonant energy build up has been proposed as a common physical mechanism behind enhancement of tunneling. When an oscillator, quantum or classical, is driven in resonance, its amplitude grows linearly with time. As a result, one of the ion's turning points gets progressively closer to a symmetrical position between two stable states, symmetrizing the polaron potential and reducing its height. After a finite number of oscillations the barrier is reduced so much that the ion tunnels ``easily.'' (Of course, tunneling is not instantaneous but because of exponential dependence, at a very high level one can assume that tunneling happens as soon as the barrier is reduced to a certain height.) After tunneling, the ion initially loses energy dropping to the potential bottom. After that, the process begins in the opposite direction. 

(iii) It follows from the above mechanism that $f_{\rm tun}$ scales linearly with the number of resonant oscillations, i.e., linearly with the potential's depth. At the same time, frequency of undriven tunneling scales exponentially with the potential's depth. Thus the ratio of driven to undriven $f_{\rm tun}$ scales {\em exponentially} with the coupling constant and as such can reach several orders of magnitude in the deep polaron regime. 

(iv) With similar reasoning, one can predict a linear dependence of $f_{\rm tun}$ on the field amplitude ${\cal E}$. (And subsequently, a square root dependence on laser intensity, $f_{\rm tun} \propto \sqrt{I}$.) The rate of amplitude increase is proportional to ${\cal E}$. Therefore, the number of oscillations needed to reach the ``easy tunneling'' condition is $\propto {\cal E}^{-1}$, from which the stated dependence follows. 

(v) In addition to affecting ions, the external field couples directly to carriers. The field modulates carrier density on both sites, which in turn rocks the ion and drives it in resonance even in the absence of direct laser-ion interaction. The two interactions are generally of the same order but they pull the ion in opposite directions, so that the two forces partially compensate each other. 

Let us conclude by briefly discussing possible consequences of polaron ``undressing''. The present work was motivated by recent experiments on dynamically stabilized superconductivity.~\cite{Hu2014} We previously proposed~\cite{Kornilovitch2016} that exponential enhancement of bipolaron tunneling rates between copper-oxygen bilayers could explain the observed increase of the apparent critical temperature in YBCO. This proposal requires additional proof. Given that the observable quantity in Ref.~[\onlinecite{Hu2014}] was an optical response, it would make sense to compute the dynamical optical response of a driven (bi)polaron. This calculation is left for future work. More generally, any polaron property that depends on the tunneling probability, for example dc conductivity or Hall coefficient, will be affected by a resonant excitation of ions. A general theory of transport properties of driven polarons seems to be worth developing.

\begin{acknowledgments}

The Author wishes to thank Alexei Voronin for help with computational resources, as well as David Dunlap, David Eagles and Vassilios Kovanis for useful discussions on the physics of driven systems.    

\end{acknowledgments}

\end{document}